# Review on modeling the societal impact of infrastructure disruptions due to disasters

Yongsheng Yang [a,b], Huan Liu [c1], Ali Mostafavi[d], Hirokazu Tatano[c]

[a] *Joint International Research Laboratory of Catastrophe Simulation and Systemic Risk Governance, Beijing Normal University, Zhuhai 519087, China.*

[b] *School of National Safety and Emergency Management, Beijing Normal University, Zhuhai 519087, China.*

[c] *Disaster Prevention Research Institute, Kyoto University, Kyoto, Japan.*

[d] *Zachry Department of Civil and Environmental Engineering, Texas A&M University, College Station, TX, United States.*

**Abstract:** Infrastructure systems play a critical role in providing essential products and services for the functioning of modern society; however, they are vulnerable to disasters and their service disruptions can cause severe societal impacts. To protect infrastructure from disasters and reduce potential impacts, great achievements have been made in modeling interdependent infrastructure systems in past decades. In recent years, scholars have gradually shifted their research focus to understanding and modeling societal impacts of disruptions considering the fact that infrastructure systems are critical because of their role in societal functioning, especially under situations of modern societies. Exploring how infrastructure disruptions impair society to enhance resilient city has become a key field of study. By comprehensively reviewing relevant studies, this paper demonstrated the definition and types of societal impact of infrastructure disruptions, and summarized the modeling approaches into four types: extended infrastructure modeling approaches, empirical approaches, agent-based approaches, and big data-driven approaches. For each approach, this paper organized relevant literature in terms of modeling ideas, advantages, and disadvantages. Furthermore, the four approaches were compared according to several criteria, including the input data, types of societal impact, and application scope. Finally, this paper illustrated the challenges and future research directions in the field.

[*] Corresponding author.

Disaster Prevention Research Institute, Kyoto University, Kyoto, Japan

*E-mail address:* huan.liu.b05@kyoto-u.jp (H. Liu)

# 1. Introduction

Infrastructure refers to assets, networks, and systems in the built environment that provide essential services (e.g., energy, water, power, transportation, and communication) for social and economic activities[1]. The terms “infrastructure systems”, “critical infrastructure”, and “lifelines” are often used interchangeably, but there are some distinctions among them. Infrastructure systems “whose reduced performance or disruption would have debilitating impacts on the defense and national security” are regarded as critical infrastructure [2]. Lifeline systems are those critical infrastructure systems that are characterized by spatially extensive network structures [1]. In the field of hazards and disasters, the term “infrastructure systems” is most commonly used; therefore, this paper uses this term throughout. In addition, given the importance of infrastructure to the safety and well-being of modern societies, different countries have defined and listed their infrastructure systems, while the following systems are consensus: energy (especially electric power), water, wastewater, transportation, and telecommunications systems.

Infrastructure systems are highly vulnerable to natural disasters, and damages to infrastructure facilities could induce a large-scale disruption of essential services. According to the World Bank report, natural disasters, such as typhoons (hurricanes), earthquakes, and floods, are a leading cause of infrastructure service disruption, and most infrastructure assets over the world are exposed to high-risk areas of natural disaster [3]. With the intensification of global climate change and physical deterioration of infrastructure, the threat of extreme hazards to infrastructure components or systems tends to be larger in the future [4,5]. In addition, infrastructure systems typically comprise geographically extensive, interdependent networks, which can improve infrastructure operational efficiency in serving large populations, but the interdependencies of infrastructure would also increase the systemic risk of infrastructure disruptions[2]. Numerous worldwide events have shown that the destruction of one infrastructure component or system can produce cascading failure, and cause disproportionately large-scale disruption of infrastructure services in multiple systems and regions [6].

Infrastructure forms the backbone of a functioning society, and the disruption of infrastructure services not only causes huge economic losses, but more importantly, it can cause significant negative societal impacts, such as the disruption of individuals' daily activities, the reduction of societal well-being, the occurrence of social panic (or even social instability), and so on[7]. Infrastructure services are essential and ingrained in modern life, for example, residents need portable water for drinking, electricity for household appliances, transportation for traveling, etc.; therefore, infrastructure disruptions could affect all aspects of people's lives, even threatening their health and survival [8]. With the continuous development of cities, a larger population becomes increasingly dependent on infrastructure services. Consequently, the societal impacts resulting from unexpected disturbances have also becomes greater [9]. Examples of negative societal impacts of infrastructure disruptions include:

- Typhoon No. 15 (Faxai) struck the Kanto region of Japan in September 2019, leaving

around 934000 and 140000 households without power and portable water, respectively. Full restoration of the power and water outage in Chiba Prefecture took about two weeks, during which more than 50% of affected households were unable to perform daily living activities such as cooking, communication, night life, bathing, and washing clothes [10].

- Hurricane Maria (Category 4) made landfall in Puerto Rico of America in September 2017, severely damaging 80% of the electrical power system through strong winds and floods, and leaving the island in a near-complete blackout. Less than 20% of the island's electricity was restored after one month, which made all communities suffered enormously from power and water outages, especially for the vulnerable groups [11].
- The Great 2008 Chinese Ice Storm occurred in the southern region of China, causing widespread power system failure, which triggered the disruption of water supply, railway system, medical service system, and supply chains, with direct economic losses of up to 156 billion yuan. Millions of people suffered from these large-scale disruptions, for instance, the disruption of the railway system was coincided with the peak of the Spring Festival (high travel demand), and about 5. 8 million people were stranded in railway stations alone, unable to return home [9]; the disruption of the supply chains led to the shortages of food and escalation of food price in 11 provinces [12].

To reduce the risk of socioeconomic impacts from infrastructure disruption, governments in different countries have developed several critical infrastructure protection plans, and researchers from various disciplines have been involved in studying infrastructure systems. The U.S. government issued the National Infrastructure Protection Plan (NIPP), which outlines how government and private sectors work together to manage risks and achieve resilience of infrastructure[13]. Similarly, Europe, Australia, Japan, China, and other countries have also made efforts to better protect their infrastructure [2,14]. This increased attention of governments attracts researchers from various backgrounds to study on modeling and improvement of infrastructure. In the last 20 years, lots of innovative and diverse work has been done on the vulnerability, reliability, and resilience analysis of infrastructure systems[15–17]. Also, the interdependency between infrastructure systems, which can lead to cascading failure propagations, has received increasing attention in the last decade [18–20]. In summary, previous academic communities put more emphases on the research of infrastructure systems themselves and contribute substantially in protecting infrastructure. Actually, infrastructure systems are critical because of their role in societal functioning, especially under situations that modern societies become increasingly dependent on infrastructure systems. However, precisely how infrastructure service disruptions impair society is poorly understood owing to the difficulties in quantitatively measuring the societal impact and integrating it with disruptions.

More recently, the academic community has also recognized the importance of exploring the societal impact of infrastructure disruption and begun to devote themselves to research in this field. For example, Hasan and Foliente [4] reviewed the literature on socioeconomic impact assessment methods of infrastructure disruption from the perspective of key stakeholders, but they mainly focused on reviewing economic impact models, which were divided into Input Output model (IO model) and Computable General Equilibrium model (CGE model). Chang [1]presented a comprehensive review of the socioeconomic impacts of infrastructure disruptions and further clarified the definition, types, measurement, and challenges of socioeconomic impacts of disruptions. Andresen et al. [21] conducted a literature review related to the social impacts of power outages in North America, and they emphasized understanding how power outages affected society and identifying the most vulnerable populations to power disruptions. Those studies provided insightful reviews on the contents and patterns of societal impacts caused by infrastructure disruptions; however, they failed to deeply review the modeling approaches of societal impact, which can be supported and complemented by the very recently cutting-edge literature.

This paper presents a review of the broad literature related to modeling the societal impact of infrastructure disruptions. To the best of our knowledge, this is the first review that comprehensively explores the societal impact modeling of infrastructure disruption from literature published over a long time. To establish a comprehensive repository of relevant literature, this paper creates a repository using a two-step search method [22]. The date range of the search and selection criteria was the period of the last three decades to avoid missing important literature related to infrastructure. The first step was "Topic" search to the Web of Science database. The topic of the relevant papers should contain at least one keyword from two categories: (1) those related to disaster and infrastructure: infrastructure resilience, infrastructure disruption, and disasters; and (2) those related to societal impacts of infrastructure disruption: societal impact, social impact, community impact, well-being impact, and socioeconomic impact. A total of 330 studies were found in the database. Then, to include influential studies that not included in the Web of Science database, the authors conducted a content-based search on Google Scholar using the same criteria and reviewed the top 10% of studies ranked by relevance for each year [22]. A total of 186 studies were selected. The authors then reviewed the titles and abstracts of the selected literature and removed irrelevant literature (e.g., societal impact studies unrelated to disaster, infrastructure disruptions, and quantitative modeling). The final number of selected papers was 113 papers. The authors then summarized the reviewed studies according to two criteria: 1) measurements of societal impacts; and 2) modeling approaches of societal impact.

The remainder of this paper is organized as follows and illustrated in **Fig. 1**. Section 2 introduces the definition and measurement of the societal impact of infrastructure, which provides the theoretical foundation for the societal impact modeling. Section 3 reviews and compares different modeling and simulation approaches. Challenges and research directions are presented and discussed in Section 4. Finally, Section 5 provides general conclusions and insights from the literature review.

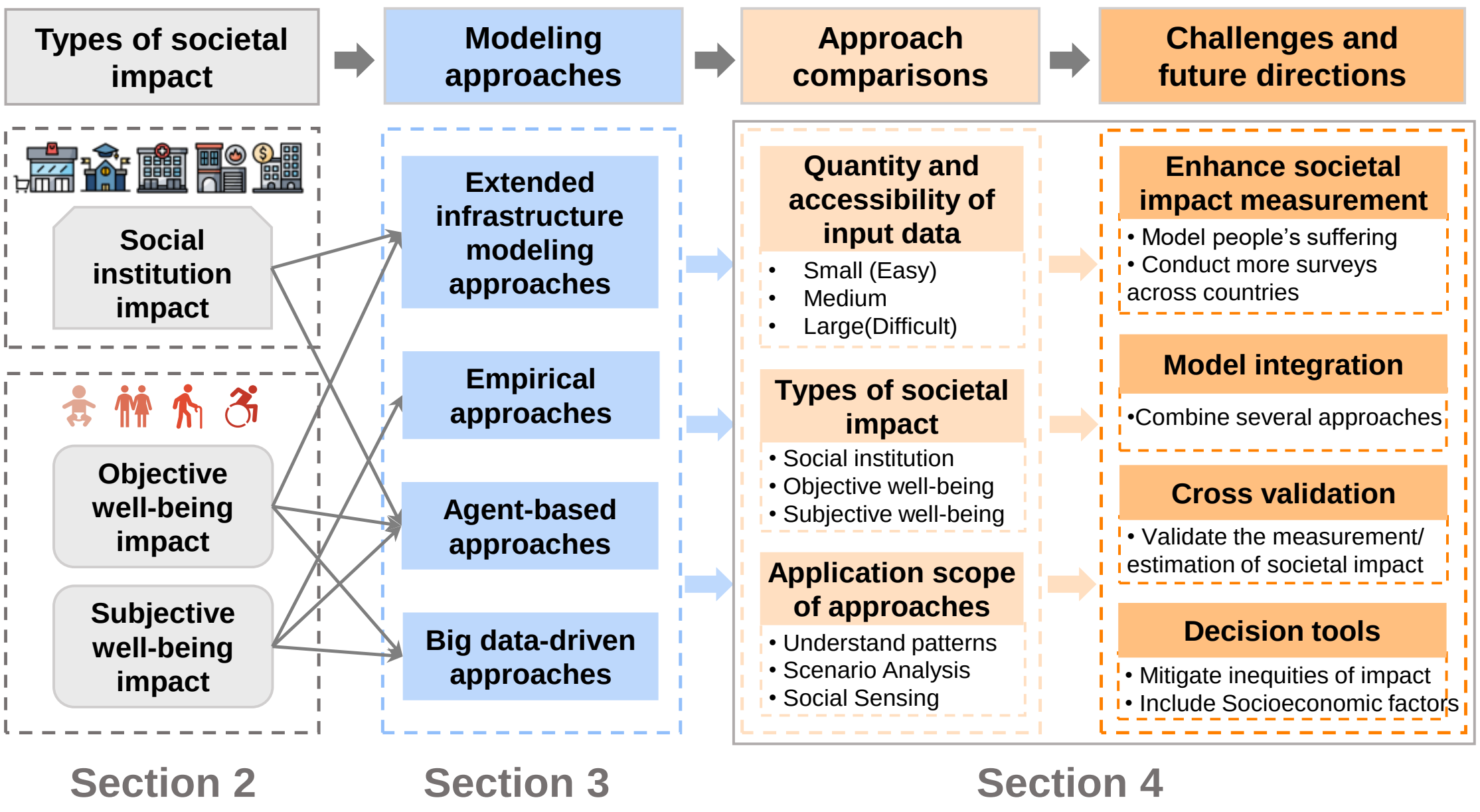

**Fig. 1** The framework of the literature review

# 2. The definition and types of societal impacts of disruptions

## 2.1 The definition of societal impact

Societal impact is the consequences of hazard-induced perturbations that can create changes in all sectors of society [23]. In a broad sense, individuals, building environments (e.g., buildings, infrastructure, facilities of factories, et.al.), institutes (medical service, emergency service, financial service, et.al.), and interactions among people all belong to the sectors of society due to the multidimensionality of society, accordingly, their changes caused by disruptive events are societal impact. When narrowed down to the field of disaster, the damages (or failures) to the built environment are usually regarded as the physical impacts, which are further quantified by the monetary losses [24]. To separate from the economic impact, the societal (or social) impact mainly refers to non-monetary outcomes of disaster on individuals, social institutes, social interactions, and public safety [7,24].

The terms 'societal impact' and 'social impact' are often used synonymously and interchangeably in the literature, though there are subtle differences between the two terms. While societal impact refers more to the impact of perturbations on various levels and sectors of society, social impact often refers to a more personal level of effects on individuals directly or indirectly [23]. Andresen et al. [21] defined the social impacts of power outages as the direct and indirect effects on people's well-being or their physical or mental health. Gardoni and Murphy [25] defined the societal impact of disruptions in terms of the impact on selected individual capabilities, the functionings individuals are able, still able, or unable to achieve in the aftermath of a hazard. At the same time, they illustrate that societal impacts should broadly include the potential effects of a hazard upon the operation of economic, social, political, and ecological systems within communities because impacts on those systems directly affect the lives of individuals within affected communities [25,26]. Social impact actually can be seen as a subset of societal impact. Given that infrastructure disruptions affect not only the individual well-being but also various social systems, this paper uses the term "societal impact" throughout the paper.

This paper defines the societal impact of infrastructure disruptions as the changes in societal functioning, which can be categorized into two groups: one is the social institution impact, and the other is the individual functioning (well-being) impact. The detailed description of types and influencing pathways of societal impact of infrastructure disruptions are illustrated in **Fig. 2** and discussed in **Section 2.2**.

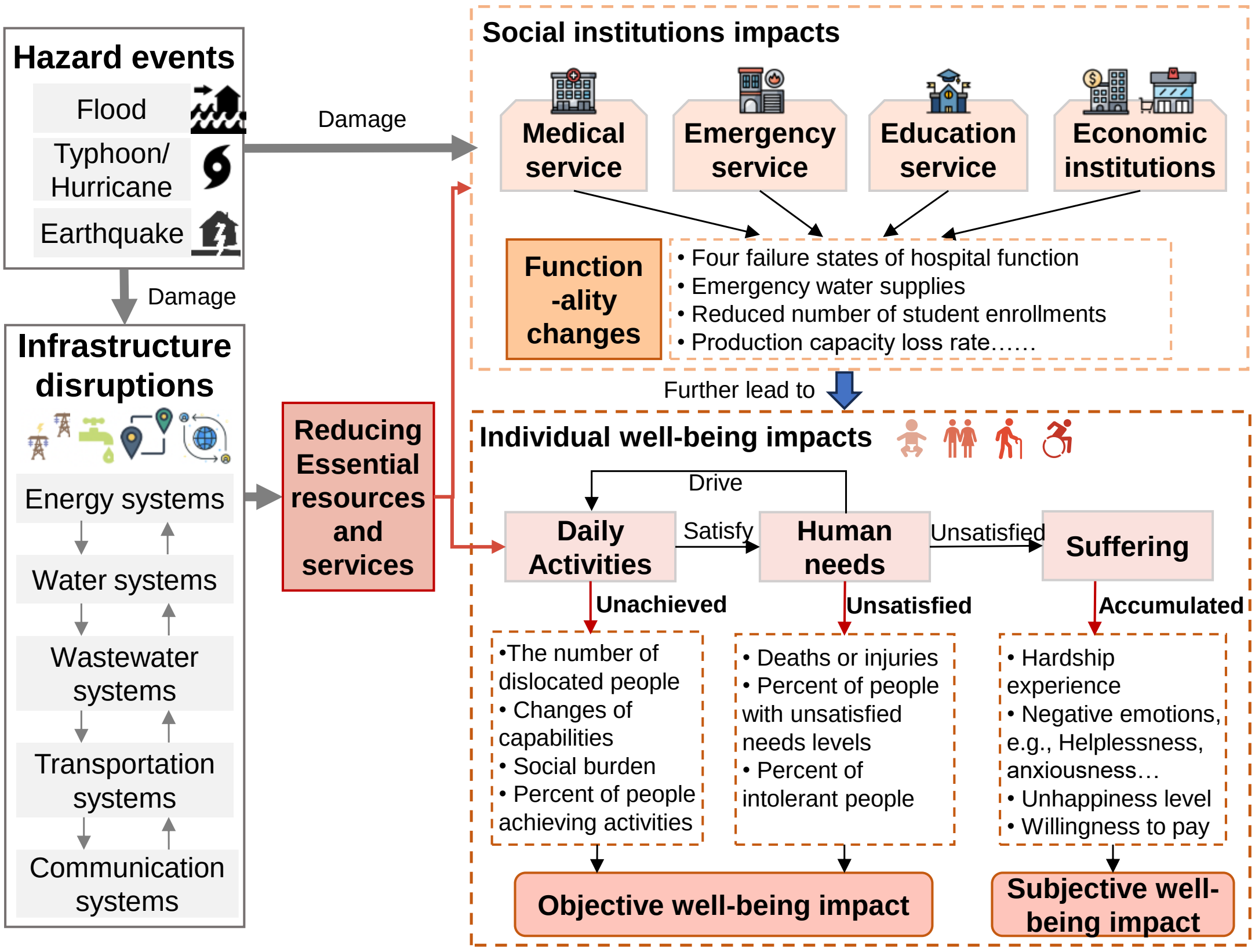

**Fig. 2** Illustration of types and influencing pathways of societal impacts of infrastructure disruptions

## 2.2 The types of societal impacts

### 2.2.1 Social institution impact

Infrastructure disruptions compromise the operations of the very institutions that the public values most highly in disaster situations, and some researchers take the change of institution functionality due to disruptions as a proxy to denote the societal impact, such as the reduced level of service of medical, emergency response, education, business, and so on.

Medical service systems (hospitals), which could provide treatments for the ill or injured people, are critical for reducing fatalities and maintaining people's well-being in the aftermath of a disruptive event [24]. Hospitals may need to curtail health care service or even shut down under disruptive scenarios of electric power, portable water, and communication; consequently, the patient and injured people may not be treated in time and people's survival is directly threatened [1]. According to the social investigation in Alameda, residents considered major hospitals to be the most important elements in the built environment under earthquake scenarios [7]. Yavari et al. [27]and Chang et al. [28] denoted the societal impact based on the reduction of health care functionality in regions considering the failures of power system, water supply system, buildings, and personnel. In details, health care functionality was assessed according to four classes: Fully functional, Functional, Affected functionality, and Not functional. Similarly, Jasiūnas et al. [29] linked the socio-economic aspects to power system disruption models, and utilized the medical service losses as one of the dimensions to represent the social impact of power system disruptions; at the same time, the number of employees in healthcare sectors without power was calculated as a proxy of healthcare service impact.

Emergency service is usually organized and conducted by governments to prevent the escalation of hazards, search/rescue people's lives, provide survival-related humanitarian relieves (e.g., food, water, temporary housing, et., al.), and the restoration of social functions. Emergency services are very dependent on various infrastructure systems, without which they can be impeded and indirectly cause the losses of human life and properties. For example, fire-fighting requires a sufficient volume of water from the water supply system, the emergency command and dispatch of resources (goods and crews) rely on the communication system, and the delivery of survival-related relieves or repair workers requires the functioning of transportation systems. Davis [30] defined the post-disaster water system service categories, which incorporated the potential impact of water disruption on society by fire protection. Yang et al. [9] measured the societal impact by considering the emergency water supplies in shelters under the disruption of transportation system and water supply systems.

Education service is the primary social institution dedicated to the transfer of knowledge, skills, and values from one individual or group to another [24]. Various national and international organizations recognize the importance of education systems to communities' stability and well-being, while education systems can be closed due to disruptions in electricity systems (support educational computers, lights, projectors, et.al.), water system (support the survival and hygiene of students or faculties), and transportation system (support traveling to schools) under disaster scenario. Hassan and Mahmoud [31] introduced the social services stability index to measure the impact of disruptive events on community and focused on healthcare and education as pivotal services which is calculated by aggregating the weight and their functionality changes over time. Aghababaei and Koliou [32] utilized the reduced functionality of education systems, specifically, the reduced number of student enrollments over time, to represent community impact given the disruption of the electric power network and water supply network subject to tornado hazards.

Businesses (economic institutions) facilitate the allocation of scarce resources across society, in mechanism, businesses produce goods and services that fulfill the multi-hierarchy needs of people, such as survival needs, career achievement needs, and social belonging needs [24]. Businesses can be disrupted by hazard events in many ways, and several surveys in disaster-affected areas indicated that lifeline service disruptions are major contributors [1,33]. Business interruptions would further cause sever socio-economic impacts, such as like lost production and sales, reduced income, and unmet people’s needs, etc. Aghababaei & Koliou [34] denoted the community impact of infrastructure disruption by functionality changes in the education system, hospital system, and businesses. They quantified the business impact by the cease operation day of businesses, unemployment rate in regions, and number of absent employees. Nozhati et al. [35] considered the effects of disrupted water supply systems, power systems and transportation systems on the functionality of commercial facilities (stores or supermarket) to evaluate food security of the society. Kajitani and Tatano [36] and Liu et al. [37] used the production capacity loss rate (PCLR) as a measurable indicator to quantify the impact of disasters on businesses and built the relationship between PCLR and disruptions of lifelines.

### 2.2.2 Individuals’ well-being impact

Individuals are the basic units that make up society, and the impact of infrastructure disruption on society can ultimately be decomposed into the impact on individuals. Substantial studies have illustrated that many aspects of individuals can be affected by infrastructure disruptions, such as their physical health, mental health, daily activities, quality of life, etc., and these impacts can be suitably covered or denoted by the well-being impact of individuals.

Well-being is a multi-dimensional concept. Disciplines define well-being in a variety of different ways, and one of the most widely cited definitions of well-being is as follows: “well-being can be understood as how people feel and how they function both on a personal and social level and how they evaluate their lives as a whole”[38]. In addition, well-being can be grouped into different categories according to the emphasis and target of different studies. For example, in terms of the domain of well-being, it usually encompasses physical health well-being and mental health well-being of individuals or society. From the perspective of measuring and analyzing well-being, it can be divided into subjective well-being and objective wellbeing. The former describes an individual’s perceptions and feelings about different aspects of their life and is measured by asking people “how satisfied are you in your…” for various aspects of their life through social surveys, such as personal health, happiness, life satisfaction, achieving in life, personal relationship and so on [39,40]. The latter (objective well-being) is concerned with measuring and analyzing the empirically observable material conditions affecting the lives of individuals [39,40]. Scholars from different disciplines usually propose some quantifiable indicators that are explained by theoretical frameworks to characterize people's living conditions, such as the human development index (income level, years of education, life expectancy, etc.) and the physical quality of life index [41]. The popular theoretical frameworks may include the capabilities theory, basic needs theory, primary goods, and so on. When focusing on the impact of infrastructure disruptions on individuals’ well-being, it can also be grouped into objective and subjective well-being impacts (**Table 1**).

Different scholars have proposed various instruments to measure the individuals' well-being impact considering influencing mechanism of infrastructure disruptions, as shown in **Fig. 2** and **Table 1**. In the dimension of objective well-being impact, researchers mainly from the engineering field proposed several indicators that are closely relevant to individuals' loss of service level (e.g., affected daily life and unsatisfied needs) to measure the well-being impact. These indicators include deaths or injuries, dislocated people, reduced capabilities, social burdens, unachieved daily activities, and unsatisfied needs. Capabilities theory [42], welfare-based approach [43], and need-based theory [9,44] are usually utilized to explain and verify the proposed indicators. In the dimension of subjective well-being impact, researchers mainly from social science proposed several indicators related to an individual's negative perceptions and feelings about infrastructure disruptions, such as the hardship experience, suffering level, and negative emotion. The theories/theoretical frameworks behind indicators are usually proposed with the validation by social survey in specific cases. Compared with subjective well-being impact, objective well-being impact has a more detailed influencing path, for example, it specified the impact transfer media (e.g., damaged house and shutdown facilities) and individuals' interruption of daily activity (e.g., drinking, cooking, bathing, and adaptive behaviors). On the contrary, more influencing factors can be incorporated into subjective well-being impact measurement attributing to the rich data in social surveys, such as the social vulnerability, household needs, emergency preparation, social capital, past experience, and so on.

**Table 1** Individuals' well-being impacts of infrastructure disruptions

| **Authors** | **Theory used** | **Indicators** | **Types of well-being impact** | **Influence path** | **Main influencing factors** |
|---|---|---|---|---|---|
| Yates [45] | N/A | Deaths or injuries | • Physical health impact<br>• Objective well-being | • Damaged infrastructure →generate forces, e.g., temperature change, pressure, electromagnetic fields, or collision → impact on individuals → casualties<br>• Infrastructure service disruption → facility shutdown →diseases , e.g., hypothermia, carbon monoxide poisoning, heart attack → casualties | Hazards from physical forces or loss of infrastructure service continuity, vulnerable people (elderly, the disabled people, poor household). |
| Masoomi et al. [46] | N/A | The number of out-migrated people (dislocate permanently) | • Objective well-being | Infrastructure disruption (electricity, water, school) → affect functionality of house, workplace (employees), and school (students) → outmigration of households | Infrastructure's functionality level and disruption duration, household states (affected houses, students, and employees) |
| Chang et al. [28]; Wang et al.[47] | N/A | • The number of dislocated people (temporarily or permanently)<br>• Population stability | • Objective well-being | Infrastructure disruption + building damages→house unhabitable→dislocation of people | Housing damage, water and electric power availability, socio-economic factors (tenure status, race and ethnicity, income level, seasonal population) |
| Gardoni and Murphy [48]; Tabandeh et al. [49] | Capability approach | • Selected 10 capabilities of individuals, e.g., Meet physiological needs, physically safety, Mobility, etc.<br>• Selected 16 indicators to represent capabilities | • Physical/mental health impact<br>• Objective well-being | Infrastructure disruption →reduce resources and services →reduce individuals' functioning (beings or doings) →well-being impact | Infrastructure's performance level and disruption duration, social attributions (age, occupation education, etc.), available resources (water truck) |
| Clark et al. [50] | Capability approach, Social burden | • Social burden metrics (defined as a function of a household's relative need to access specific services divided by household's accessibility to those services) | • Physical/mental health impact<br>• Objective well-being | Infrastructure disruption →reduce resources and services →take adaptive measures to fulfill needs→ reduce individuals' functioning (beings or doings) →well-being impact | Relative need of population to achieve capability types, accessibility to service-providing locations (total travel costs, total direct costs, opportunity costs) |
| Yang et al. | Maslow's | • Percentage of people at five | • Physical health | Infrastructure disruption → reduce resources and | Water quantity in the whole society: |
| [9] | hierarchy of needs | need satisfaction levels | impact<br>• Objective well-being | services→ disrupt individuals' daily activities→ unsatisfaction of essential service needs→ societal impact | considering the availability of tap water, bottled water, emergency water. |
| Yang et al. [51] | Capability approach, activity choice | • Percentage of people achieving certain activities<br>• Percentage of people get intolerant | • Physical/mental health impact<br>• Objective well-being | Infrastructure disruption → reduce resources and services→ disrupt individuals' daily activities→ become intolerant→negative well-being | Socio-demographic factors (age, car, household size, etc.), types and duration of infrastructure disruption, emergency preparation and resources, past experience |
| Silva-Lopez et al. [43] | Welfare economics | • The expected welfare loss per commuter | • Mental health impact<br>• Objective well-being | Road network disruption→increase commute time→ welfare losses | Disruption of road networks, different income groups (low, middle, high) |
| Dhakal and Zhang [52] | Welfare economics | • Gini coefficients that measure unequal distributions of functional loss and recovery time | • Objective well-being | Infrastructure disruption → percent of infrastructure disruption in regions → Time required to recover in regions → unequal impact | Functional loss of infrastructure in regions, recovery time in regions |
| Esmalian et al. [53] | Hardship experience | Percent of households having hardship experience (outage duration is larger than tolerance time) | • Physical health impact<br>• Subjective well-being | Infrastructure disruption → disruption duration exceeds the tolerable time→ hardship experience→ well-being impact | Disruption duration, social vulnerability (race, inhabitant time), household needs, emergency preparation, availability of substitute, social capital, past experience |
| Dargin and Mostafavi [54] | Personal wellbeing index | Helplessness, anxiousness, upsetting thoughts, depression, and so on (measured in five-point Likert-scale from 1 to 5) | • Mental health impact<br>• Subjective well-being | Infrastructure disruption→ affect different dimensions of well-being | Socio-demographic factors (age, race, income, etc.), types and duration of infrastructure disruption, emergency preparation, past experience |
| Stock et al. [55] | Proposed a conceptual framework | • Unhappiness level (from 0 to 1)<br>• Willingness to pay (WTP, $) | • Physical/Mental health impact<br>• Subjective well-being | Infrastructure disruption → disruption duration → increase unhappiness level and WTP →well-being impact | Disruption duration, socio-demographic factors (age, income, education, members with medical condition), risk perception, past experiences, emergency preparation, social capital |

# 3. The models for societal impact evaluation

This section reviews the existing modeling approach for societal impact estimation of infrastructure disruptions. They are broadly categorized into 4 groups: extended physical infrastructure modeling approaches, empirical approaches, agent-based approaches, and big data-driven approaches. The inputs, connecting methods, outputs, strengths, and weaknesses of each type of approach are summarized in **Fig. 3**, and details are illustrated in the following subsections.

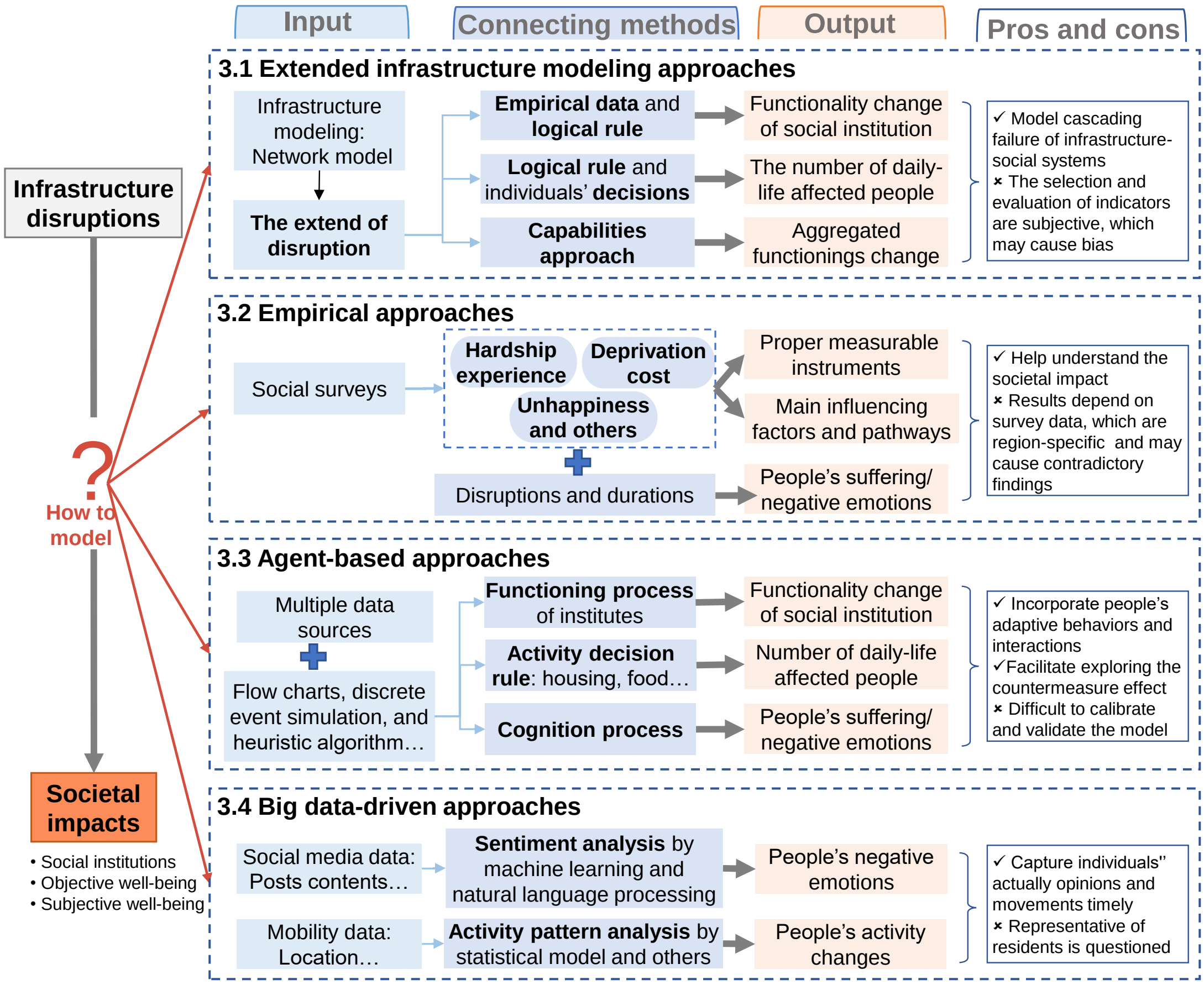

**Fig. 3** Overview of four types of societal impact modeling approaches

## 3.1 Extended infrastructure modeling approaches

Extended physical infrastructure modeling approaches estimate the societal impacts by integrating the physical failure analysis of infrastructure systems (engineering dimension) and change analysis of social systems (social dimension). In engineering dimension analysis, the functionality and interdependency of infrastructure systems are modeled in ways that support estimating the societal impact. Given the social institutions and individuals are located in different spatial regions, network-based models are usually adopted to calculate the spatial distribution of infrastructure service disruption. In social dimension analysis, the susceptible sectors of society to infrastructure disruption are identified and quantified by some indicators. More importantly, the relationship between disruptions and selected indicators is established to derive the societal impact. This type of approach advances in connecting the failed infrastructure components with the societal impacts and enables modeling the cascading failure of interdependent infrastructure-social systems. This approach is widely used in community resilience or infrastructure resilience assessment that includes societal consideration; usually, the societal impacts are quantified by social institution impact and individuals' objective well-being impact.

### 3.1.1 Extending method for social institution impact estimation

In aspects of social institution impact, the functionality of social institutions is generally quantified and estimated by modeling the relationship between institution functionality and infrastructure disruption, and their relationship is mainly established by empirical data and logical rules. For example, Chang et al. [28]quantified healthcare facility functionality considering lifeline disruptions based on damage data from post-earthquake safety inspections of 228 facilities, and the functionality class probabilities of healthcare facility would be adjusted up one level, if it experienced loss of at least one external lifeline. Liu et al. [33] modeled the dependencies of production capacity on lifeline disruptions in different business sectors using production functions that were fitted using a dataset from a post-disaster business survey for the 2011 Great East Japan Earthquake. While in most cases, the damage data are not fully recorded or even not available, scholars established different logical rules to connect infrastructure disruption to institutions' functionality. Jasiūnas et al. [29] developed an integrated spatial rule for linking disruptions in a power system with critical service (healthcare service), which were represented and calculated by accumulating the share of the disrupted power supply through time and space, e.g., the average time that number of people employed in health sectors without power. Loggins et al. [56] modeled the interdependencies between civil infrastructure and social infrastructure (e.g., the police and fire services, healthcare services, critical commercial services, etc.) based on setting constraints that denote different effects of civil infrastructure on demand, supply, or transshipment nodes in social infrastructure. Hassan and Mahmoud [57]established the relationship between a hospital's functionality and the disruption of infrastructure using success tree, where AND/OR gates are used to connect the basic events (infrastructure) to top events (hospitals) and intermediate events. A similar methodology has also been proposed to model the functionality of education systems [58].

### 3.1.2 Extending method for individuals' objective well-being impact estimation

In studies related to individuals' objective well-being impact estimation, two types of methodology are popularly adopted by scholars. One is identifying one or several indicators that can mostly represent the individuals' well-being impact of infrastructure disruptions by expert judgment or practical experience, and then separately evaluating it/them by mapping the disruption to the affected population. The other methodology is applying the theory/approach used to measure the objective well-being of individuals in social science in the evaluation of infrastructure disruption (e.g., the capabilities approach is the most popular theory to capture the well-being impact), and accordingly, the connection algorithm between infrastructure disruptions and well-being impact are developed.

**(1) Extending infrastructure disruption to the affected populations**

In Method (1), the number of populations without life-related services is usually calculated to indicate the objective well-being impact, and logical rules and households/individuals' decision-making processes are designed to extend the infrastructure disruption to societal impact. For example, Nozhati et al. [35] measured the well-being impact of disruptions by the number of food-secure people, which is estimated by the number of people who can access the functioning stores under disruptive events, and food retailer is functioning only if its building structure, water, and electricity are available. Yang et al. [9] defined the societal impact of water suspension as the percentage of the population in each need satisfaction level, which depended on the available water quantity in disasters, and the availability of tap water (water supply system), bottled water (commercial stores), emergency water (equal distribution rule) are modeled and integrated into each spatial population grid. Masoomi et al. [46] quantified the socioeconomic impact by population outmigration, the probability of which depends on state changes of households (affected houses, affected students, affected employees) due to disruption and recovery of physical networks (i.e., electric power network, water network, and buildings). The above connecting rules are determined by the way that infrastructure disruption affects people's daily lives, and they simplify individuals' decision process (e.g., going to the nearest store, getting water, out-migrating, etc.), which are the key to connect disruption with well-being impact and are affected by multiple factors.

Individuals' response or adaptive decisions under infrastructure disruption will determine the availability of life-related services and further contribute to the well-being impact. Scholars have proposed a logic-tree approach, discrete choice modeling, and optimization approach to model individuals' decisions. Chang et al. [28] modeled the social impact of lifeline losses by displaced persons, and it was evaluated by a logic-tree type of approach, which simulated the households' decision-making process considering housing damage, lifeline loss, socio-economic and locational factors (car ownership, elderly, ethnicity, tenure, etc.). The number of displaced people is a popular indicator of social impact (social instability) caused by disaster [47,59]. Lin [60] developed a dislocation choice model based on a logistic regression model, which is capable of estimating the probability of households choosing dislocation considering residential structural damage and multiple socio-economic factors. Based on this model and Bayes' theorem, Beck and Cha [61] estimated the dislocation probability and expected dislocation population given the power outage probability of each node due to hurricane damage. Similarly, Nofal et al. [62] integrated the dislocation model with Housing Unit Allocation (HUA) method considering inaccessibility of transportation system and power system. This model could provide a dislocation probability for each housing unit and aggregated dislocated population. In addition, Yang et al.[51] measured the societal impact of water disruption by the number of people who can perform certain activities and the number of people who get intolerant due to disrupted activities, which are calculated using an individual's activity estimation model driven by minimizing the suffering of people's disrupted activities under limited water.

In general, this method quantifies the well-being impact of infrastructure disruptions by the number of daily-life affected people, which can be intuitive and easily quantified without complex aggregation or other transformations. However, these indicators were usually proposed by scholars' practical experience, lacking of underlined theoretical basis verifying them. The relationship between affected activities and well-being impact is not clarified. In addition, logical rules and individual decision-making models are developed directly to connect the infrastructure service disruption to societal impact; thus, it is suitable to conduct scenario analyses and answer "what-if" questions by combining physical infrastructure models. It is worth noting that the spatial distributions of infrastructure disruption are usually required to derive the affected population because, in different spatial disrupted regions, households with different socio-economic characteristics may suffer inequitable impact. The heterogeneity of populations should be considered, while the population characteristics are statistically recorded in a relatively large scale, like the level of census or grids. Recently, the HUA method has been established to link the detailed household characteristics to a spatial inventory of residential housing structures, which further narrows down the scale of the evaluation. Correspondingly, to generate the refined spatial distribution of disruptions, network-based approaches are most popularly utilized in societal impact estimation [1,63].

**(2) Integrating infrastructure disruption with capabilities-based approach**

Many theories or approaches are proposed by social scientists to measure the objective well-being of humans, where the capability approach is widely applied and integrated into the field of disaster or infrastructure disruptions. The capability approach facilitates to answer two main questions: the first one is what is the true essence of the objective well-being of individuals, including their measurement and relationship with daily-life activities fulfillment (or other critical factors); the second one is how to quantitatively estimate the well-being impact of individuals under disturbances.

The capability approach was first introduced by Amartya Sen in the development of economics to gauge the well-being or quality of life of individuals as a way of determining the overall level of development of societies[64,65]. The approach emphasizes that the well-being of individuals depends on their capabilities to lead a life that they consider valuable. To define the capabilities, they first introduced the concepts of the functioning of individuals, which refer to doings(activities) and beings (states) that individuals find valuable to do or achieve. Doings (or activities) may include eating, drinking, going to the hospital, working, etc., and beings (or states) may include staying healthy, staying safe, staying happy, etc. Capabilities thus describe the genuine opportunities or freedom open to individuals to achieve functioning (activities and states), depending on the individuals' available resources, characteristics, and social and environmental conversion factors [66]. An individual's capability is a collection of functioning under certain conditions; the greater the individual's capability, meaning that more activities in the collection can be achieved, the greater the freedom of choice of life (functionings) available to the individual, and further the greater the well-being.

Murphy and Gardoni [67] first applied the capability approach to the field of disasters in 2006 and defined and gauged the societal impact of disasters in terms of changes in individuals' capabilities. They pointed out that disasters can directly or indirectly damage individuals' living conditions and reduce their available resources, leading to reductions in individuals' opportunities to achieve functionings, i.e., some daily activities and states of individuals are inadequately achieved or even disrupted, which further reduces people's well-being [42]. In the last decade, the application of the capability approach in the disaster community has become increasingly developed, evolving from an initial theoretical framework, to qualitative evaluation methods, and to quantitative evaluation methods applied to infrastructure disruptions.

In the quantitative estimation of the well-being impact of infrastructure disruptions, scholars proposed a four-step indicator-based method founded on the capability approach. In general, the method consists of four main steps [48]: 1) Selection of the capabilities of individuals, this step identifies the specific functionings (e.g., drinking, eating, traveling) that are critical and likely affected by infrastructure disruptions. 2) Selection of indicators, because capabilities are not directly measurable, indicators for given functioning are selected as proxies, such as the frequency of drinking water supply problems, frequency of food supply problems, travel time to the nearest store, etc. 3) Developing various models to predict indicator values, taking into account the disruption of infrastructure, characteristics of individuals, damages of buildings, and losses of other living conditions. In this step, regression models and infrastructure network models are usually constructed using available data. 4) Establishing aggregation algorithm for indicators' values to represent the whole well-being impact of individuals, and evaluating the levels of well-being impact due to disruptive events.

Based on the general quantitative estimation method, scholars from different backgrounds continue evolving and improving the algorithm of each step. Steps 1) and 2) heavily rely on expert experience, literature review, or qualitative analysis (examples are shown in **Fig. 4**), which are the main research topic of social scientists [68]. Step 3) is the key to connecting the infrastructure disruptions with societal impacts, and engineers usually put more emphasis on this part. For example, Tabandeh et al. [49] and Wang et al. [68] developed a probabilistic prediction model and multinomial logit regression model for indicator indices of functioning using social survey data, which takes into account main influences factors, such as the status of infrastructure systems, personal characteristics, and resources. It is worth noting that the service statuses of infrastructure are usually simulated or predicted under disaster scenarios through developing physical infrastructure models, e.g., network-based mode [69,70]. As for the aggregation of indicators in step 4), Tabandeh et al. [71] proposed a reliability-based methodology that describes personal well-being as a series system consisting of different functioning indicators, where a "failure" in any of the functionings (the indicator value of activities is below a certain threshold) can lead to a "failure" of the individual well-being system (the individual becomes intolerable). To determine the “failure” threshold, Murphy and Gardoni [72] defined three states for indicator indices with the same labeling as the capability states, i.e., acceptable, tolerable, and intolerable, as shown in **Fig. 4**. When the level of activity achievement (indicator value) exceeds the acceptable threshold, the individual is acceptable. A state below the acceptable threshold is tolerable if its achievement is temporary and reversible, and above a minimum tolerability threshold. The tolerability threshold is the absolute minimum level of activity achievement below which the individual becomes intolerable. Besides considering the achievement of functioning, Tabandeh et al. [49] incorporated the time dimension in evaluating the indicator indices; for example, the tolerable state of the indicator indices would become intolerable if the required recovery time to improve to the acceptable state exceeds a reference duration.

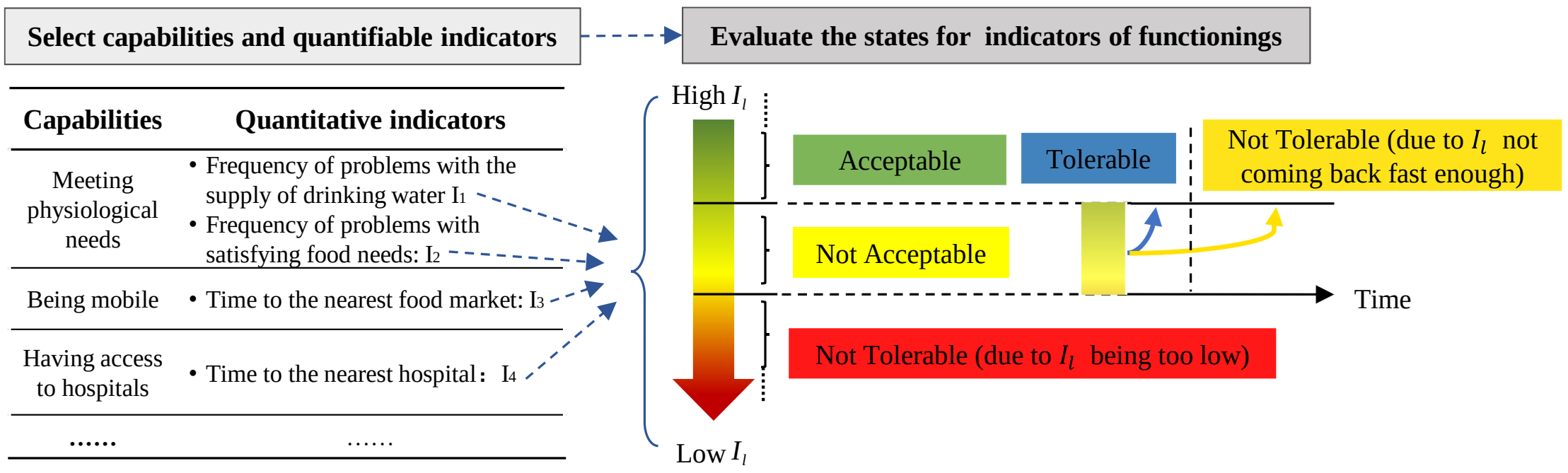

**Fig. 4** Selection of capabilities/indicators and their evaluation threshold [49]

In summary, Method (2) integrates infrastructure disruption with a capabilities-based approach, which is an effective attempt to connect engineering dimension analysis of infrastructure with social dimension analysis. This method advances in providing theoretical foundations for the measurement and estimation of the objective well-being impact of infrastructure disruptions, and it helps to clarify the relationship between the achievement of activities (functionings) and the well-being impact of individuals. Nevertheless, in the quantitative estimation methodology, the selection of indicators for functionings (Step 2) is controversial in terms of whether they are representative or effective. Additionally, in the evaluation of indicator indices (Step 4), the acceptable and tolerable thresholds are usually subjectively determined, which could cause a large bias on the evaluation results of well-being impact. Regarding this deficiency, scholars proposed the concept of tolerable level (time) of households quantified by social survey in subjective well-being impact estimation, which will be further discussed in Section 3.2.

## 3.2 Empirical approaches

Empirical approaches quantify the societal impact of infrastructure disruptions according to historical disaster data and social surveys of individuals affected by actual disasters, as well as surveys with hypothetical disaster scenarios. This type of approach can develop an understanding of what happened in infrastructure disruptions, or what could happen to disaggregate units, such as individual people or households, businesses, and organizations; and their corresponding impacts are usually quantified by individuals' subjective well-being impact. Inequitable impacts of infrastructure disruption to vulnerable populations and spaces are mostly highlighted, and their potential influencing factors are identified and understood to promote human-centered infrastructure resilience. In addition, based on empirical data, several vulnerability models for individuals or households can be developed and embedded into infrastructure or agent-based models to estimate the societal impact under various infrastructure disruption scenarios.

### 3.2.1 Quantification of well-being impact and identification of its main influencing factors

To understand what happened to people under infrastructure disruptions, appropriate instruments should be proposed to quantify the well-being impact before surveying individuals or households. As mentioned in Session 2, considering the influencing features of infrastructure service disruption and measurability of people's feelings or perceptions, scholars proposed several metrics to measure the individuals' well-being impact, which can be categorized into (1) hardship experience, (2) deprivation cost, (3) unhappiness, and other emotional well-being impact.

**(1) Hardship experience**

Quantifying the well-being impact of service disruption by the hardship experience of households/individuals is intuitive, and the tolerance-level-based method is widely developed by scholars to indirectly measure the hardship experience of households or individuals. The tolerance-level-based method was first proposed by Esmalian et al. [53] under the capability theoretical framework, and tolerance level refers to the maximum amount of time that a household or an individual can tolerate service disruption in disasters. They illustrate that hardship experience is a function of the difference between the duration of infrastructure disruptions and the household's tolerance level (tolerable time). The smaller the difference, the greater the people's suffering level; when the duration of infrastructure disruption exceeds the tolerance level, people have hardship experience, resulting in negative well-being [73,74].

The key to this method is the introduction and quantification of the tolerance level, which indicates an individual's or a household's ability to cope with and withstand the disruption[53] . The tolerance level is mainly obtained by surveying individuals about the maximum number of days they can tolerate different infrastructure disruptions (e.g., power, water, transportation, etc.). The tolerance-level-based method improved the application of the capability approach in the field of disaster and facilitated understanding of the threshold of individuals' functionings (Tolerability threshold) due to disruption. However, this method did not directly measure the hardship of individuals, instead, it regarded the hardship experience as a Boolean variable, meaning that people experienced hardship when the duration of disruption exceeded the tolerance level.

**(2) Deprivation cost**

In addition to introducing a tolerance level to measure the hardship, scholars from the field of humanitarian relief proposed deprivation cost to directly measure individuals' suffering level due to lack of life-supporting resources, such as water, food, medical service, and sanitation supplies, where many of shortages are caused by infrastructure disruptions. Deprivation cost is calculated by the economic cost, and it was initially proposed to optimize the distribution of relief by minimizing the society's suffering level. Holguín-Veras et al. [75] first proposed the concept of deprivation cost, and summarized the general characteristics of individuals' deprivation cost function as follows:

(a) Individuals' suffering level exhibits monotonically increasing, nonlinear, and convex functions with the duration of disruption increasing, as shown in **Fig. 5;** these properties reflect the body's natural response to deal with a shortage of life-supporting resources. For instance, at first, most healthy individuals can handle short-term resource disruptions, as the body's reserves of resources are used up, people's suffering level surges rapidly until it reaches a maximum value (death).

(b) Individuals' suffering level has a non-cumulative nature of demand for resources. Considering human physiology, the required amount of resources is not cumulative as the duration of the disruption increases, e.g., when an individual has no food for three days, his/her demand for food on the 4th day is limited and not the sum of the previous three days.

(c) Hysteretic effects of suffering level may exist after needs are satisfied. When an individual suffers a lot from shortages and causes irreversible damage (health impairment) to the body, individual's suffering level cannot return to its initial value after his/her needs are satisfied and creates a residual impact. Conversely, non-hysteretic effects mean that an individual's suffering and body damage can be recovered to its initial value after his/her needs are satisfied.

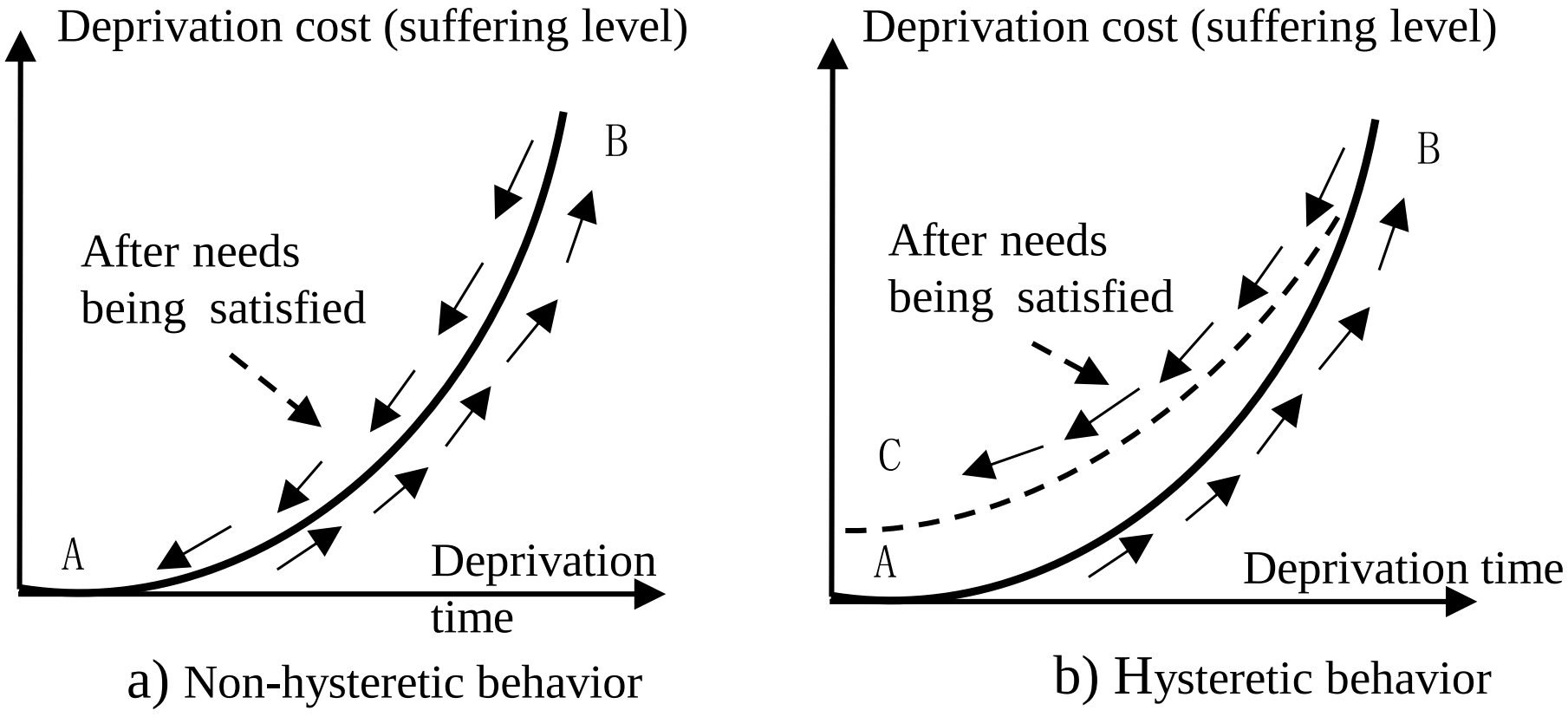

**Fig. 5** Schematic of deprivation cost function[75]

To derive the deprivation cost function, scholars developed economic evaluation methods, such as the Contingent Valuation, Conjoint Analysis, and Stated Choice method, which assign monetary values to non-tradable goods or services (e.g., suffering level). Holguín-Veras et al. [76] applied the Contingent Valuation method to evaluate the economic costs of individuals' suffering level under water suspensions, and based on a social questionnaire, they investigated people's Willingness To Pay (WTP) to improve the situation or buy a substitute resources in a hypothetical disruption scenario. Also, the limit value was considered in the deprivation cost function, e.g., the point at which an individual dies after five days of water deprivation, and by regression analysis, an exponential function was fitted the best for the deprivation cost function. Similarly, Stock et al. [55] measured the societal impact by households' WTP to avoid electricity and water disruptions. Although they did not introduce the deprivation cost function, they also found the nonlinear relationship between infrastructure (outage duration) and societal functioning (WTP) using survey-based data from Los Angeles County, USA. Macea et al. [77] used discrete choice modeling to establish a deprivation cost function of water disruption based on a social survey with questions about various hypothetical disruption scenarios, and they found that the Box-Cox model fitted the function best. Since the above studies did not consider individual heterogeneity, Macea et al. [78] incorporate more factors into the deprivation cost function based on a discrete choice model, such as individual attributes, risk perceptions, safety culture, and trust.

**(3) Unhappiness and other subjective well-being impact**

Dimensionless scales in social science or psychological science can also be adopted to measure individuals' physical and cognitive feelings (subjective well-being impact) about infrastructure disruption, for example, the 5-point Likert scales, 11-point numerical rating scale, and customized rating scales. Dargin and Mostafavi [54] utilized 5-point Likert scales, ranging from None at all (= 1) to A great deal (= 5), to measure people's subjective well-being impact of infrastructure disruption by hardship experience and emotion changes, like helplessness, anxiousness, upsetting thoughts, and depression. These impacts are derived by surveying the affected households and asking them: “What was the extent of overall hardship/emotional impact experienced due to lifeline outages/interruptions posed by disasters” . Wang et al. [79] introduced a numerical rating scale (11 points) from the field of medical science to measure the suffering level due to the shortage of food, medicine, and tent during disasters, and similarly, asked people through social surveys about their degree of suffering when faced with different scenarios (0 implies no suffering, 10 implies extreme suffering). Stock et al. [55] developed two empirical measures of societal impacts: a WTP to avoid lifeline service interruptions and a constructed scale of unhappiness, which has 5 levels of unhappiness (from Not unhappy to Extremely unhappy) allowing individuals to choose in a questionnaire. They found that unhappiness is better able to distinguish the effects of shorter-duration outages than WTP is. This type of method has advantages in understanding who suffers, how much is the subjective impact, and what is their main influencing factors. However, in these studies, the measurement of subjective well-being impact is dimensionless and relative, indicating that it is difficult to develop mechanism-based model to estimate the impact, as a result, this method is rarely to be used to conduct scenario analysis of infrastructure disruptions.

After determining the above instruments to quantify well-being impact, their corresponding main influencing factors can be further explored by various statistical methods, such as correlation analysis, ANOVA analysis, structural equation modeling and regression model. For example, Esmalian et al. [53] implemented a Poisson regression model to account for the simultaneous effect of multiple factors, and they found that households’ need for utility service, preparedness level, the existence of substitutes, possession of social capital, past experience, risk communication, race and residence type mainly influence the tolerance level, and hence the level of hardship experienced in the context of the 2017 Hurricane Harvey. With the same dataset, Coleman et al. [73] adopted Spearman bivariate correlation analysis to understand the association of sociodemographic characteristics with the hardship experienced and tolerance level, and they concluded that certain socially vulnerable groups (low income, racial minority, and younger residents) reported significant disparity in the hardship experience. The same results were presented by Dargin and Mostafavi [80], and they also found that disruptions in transportation, solid waste, food, and water infrastructure services caused more significant well-being impact disparities. By applying ANOVA one-way tests and structural equation model, Dargin et al. [54] found that physical attributes of community, preparation behaviors, and the coupled durations of infrastructure disruptions were significantly associated with household hardship experience. Households’ poor preparation is attributed to past experiences and social vulnerability, which refers to the households with children, racial minority status, and low income and educational attainment, and they are prone to underestimate the impacts of a disaster, or have greater barriers to preparing for disasters, such as relatively high costs, lower accessibility to stores, lower availability of store supplies, etc. Stock et al. [55] used survey-based data from Los Angeles County, and also found significant role of preparation and durations of disruptions (power and water supply system) on households' well-being impact. Differently, in Stock's study, the effects of some sociodemographic characteristics were not significant, like gender, race, education level, and the household with children, partially conflicting with Dargin's findings. Furthermore, Coleman et al. [81] collected survey data from Hurricane Harvey, Hurricane Florence, and Hurricane Michael, and explored the main influencing factors of the level of susceptibility for households by Spearman correlation analysis, respectively. They highlight that some variation in the influence of factors was event-specific or service-specific, but without doubt, certain influencing characteristics have a universal impact on the well-being impact of households, e.g., households with low socioeconomic status.

In general, the current researches establish a fundamental empirical basis for understanding the households'/individuals' susceptibility and well-being impact of infrastructure service disruptions by identifying measurable instruments and their corresponding influencing factors. They facilitate the development of mathematical models based on the identified influencing factors to determine societal impact. However, empirical studies heavily depend on the questionnaires and the backgrounds of respondents, and their results are usually region-specific or hazard-specific, which could lead to contradictory findings. The current consensus that can be reached is the existence of social inequality in the well-being impact of infrastructure disruption, and socially vulnerable groups are disproportionally affected under the same degree of disruptions.

### 3.2.2 Empirical-based impact evaluation

According to survey data related to individuals' well-being impact, an empirical-based impact evaluation model can be developed to identify susceptibility for households and to connect infrastructure service disruption with societal impact. The tolerance-level-based method has advantages in modeling people's susceptibility, because it represents households' capability to withstand disruptions (tolerable days) and it can be easily transferred to people's hardship experience (negative well-being) by comparing the duration/recovery of the outage. Scholars from different countries modeled the tolerance level of infrastructure disruptions using empirical survey data. For example, based on empirical survey data collected after Hurricane Harvey, Esmalian et al. [74] established a negative binomial regression model to predict the tolerance level of households considering household characteristics (e.g., sociodemographic, social capital, resources, and previous disaster experience). By integrating this model with existing models for power outages and service restoration under an agent-based framework, the societal impact can be estimated. With the same dataset, Dong et al. [82] proposed a Disruption Tolerance Index (DTI) for healthcare service disruption using principal component analysis, and combining DTI with the community's physical vulnerability to access to healthcare facilities (transportation disruption), they found the hotspots and cold spots of the physically/ socially vulnerable communities. Based on post-disaster social survey data from Hurricane Harvey (2017),

Hurricane Florence (2018), and Hurricane Michael (2018), analogous to fragility curves for engineered systems, Esmalian et al. [83] developed susceptibility curves for disruptions in eight infrastructure systems using survival analysis models, and found that the proportion of households having hardship experience becomes larger as the duration of the disruption increases. In addition to empirical studies in the United States, Petersen et al. [84] conducted a social survey in the city of Barreiro, Spain, focusing on the analysis of water disruptions, and examined the change pattern of tolerable population over time under different influencing factors. In a Japan case study, Gentaro et al. [85] modeled the probability distribution of tolerance levels for water-related activities by conducting a social survey in Osaka, Japan, and explored the proportion of the population that becomes intolerable with increasing disruption duration. They found that tolerance level for disrupted cooking and toileting corresponded to the lognormal distribution, while disrupted bathing and laundry followed a Weibull distribution.

In terms of empirical modeling for deprivation cost function, as mentioned before, scholars from the field of humanitarian relief adopted economic evaluation methods to determine suitable functional forms, such as exponential function, Box-Cox regression model, logistic function, etc. Yang et al. [51] introduced the concept and application of deprivation cost function to the field of infrastructure resilience for the first time. Given that both hardship experience and deprivation cost represent individuals' suffering, they integrated the tolerance level into the deprivation cost function to derive the suffering level function for disrupted activities due to infrastructure disruption. Utilizing the revised suffering level function, they proposed individuals' decision-making model to assessed the well-being impact of water infrastructure disruptions. As for the unhappiness modeling, Stock et al. [55] fitted an ordinal logit with mixed effects to predict probability of household at least each unhappiness level, as a function of infrastructure type, outage duration, and household attributes. Dulam and Davidson [86] applied this model into the case study of 1994 Northridge earthquake, and estimated the spatial distribution of unhappy people.

Overall, Empirical-based impact evaluation model estimates the well-being impact of people using statistical model based on survey data. In current studies, the tolerance level is usually modeled and compared with outage duration of infrastructure to further derive households' susceptibility or well-being impact. At the same time, the empirical study with tolerance level can improve the evaluation threshold of functioning in the Capability approach. However, this type of method, as mentioned before, largely depends on survey data and characteristics of society and hazards, and it can only provide accurate estimation for future similar hazards or investigated places within the range collected dataset. Empirical studies among different countries and various hazards need to be strengthened and compared to develop widely accepted statistical models. Also, other modeling and simulation approaches should be built and incorporated for cross-validation or additional decision support.

## 3.3 Agent-based approaches

The societal impact of infrastructure disruption is a dynamically complex process, and various coupled factors influence the negative impact, such as households' and governments' protective behaviors, social vulnerability attributes, availability of emergency resources, etc. To incorporate these influencing factors and simulate the societal impact of disruptions, one effective approach is agent-based modeling, which is a bottom-up method for simulating the complex system by designing multiple individually autonomous agents and setting up their decision-making and interaction rules. Agent-based modeling for societal impact is advanced in: 1) considering the heterogeneity of agents (social vulnerability attributes) and modeling interactions among agents and environments; 2) simulating agents' nonlinear decision-making behaviors (adaptive behaviors); 3) dealing with situations where data collection and experimentation are difficult; and 4) allowing for rapid evaluation of policies/measures and the incorporation of stochastic disturbance [87,88]. Based on these strengths, agent-based approaches are increasingly used to simulate the impact of disruption on social institute function and individuals' well-being with consideration to multi-agents adaptive behaviors.

### 3.3.1 Agent-based modeling for social institution impact

Agent-based modeling has been highlighted in the field of disaster for many years, especially for flood risk management [89,90]. Scholars incorporate various individuals' or emergency agency's adaptive behaviors into disaster impact models to better estimate the disaster risk or impact considering social responses, and more importantly, to explore the effective measures to reduce the impact or risk of disasters. The adaptive behaviors may include but are not limited to: individual agent's evacuation, emergency preparation, social mutual help, and buying insurance, and decision-maker agent's reinforcement of engineering structures, early warning, emergency responses, and recovery strategies [89,91]. Similarly, in the topic of interdependent infrastructure modeling, the agent-based approach is also proved to be a powerful tool to account for various types of dependency, especially facilitating the modeling of the interdependency of infrastructure systems and social systems, which are pointed as the future directions by several related review papers [2,4]. After about one decade of development, noticeable progress has been made in integrating disruptions of interdependent infrastructure with social systems using agent-based frameworks. It is worth noting that, different from the extended infrastructure modeling (Section 1), the agent-based approaches put more emphasis on social dimension analysis. Specifically, they focus on simulating the detailed functioning process of social institutions, dynamic behaviors of individuals/households, and their interactions under infrastructure service disruptions.

In terms of agent-based modeling for social institution impact of infrastructure disruptions, the institution is usually treated as the main agent, within which the functioning process under disruptions is simulated by setting decision and interaction rules. The decision rules can be designed by flow charts, discrete event simulation, and heuristic algorithm, at the same time, the interaction rules with infrastructure and household/individuals are always highlighted. For example, Aghababaei and Koliou [32] proposed a comprehensive agent-based model for education systems, and the model incorporates the behaviors and interactions of multiple agents: schools, households, power systems, water systems, and construction companies. The decision-making process of school agent is designed by a flow chart about whether to distribute students to other operational schools according to the damage and recovery of buildings and lifelines under hurricane disasters, which further affect the school status of students of household agents. Also, households may move to other places for housing after damages, reducing the student enrollments in the education system, and the housing status of household agents is simulated by Markov chain model with consideration to socio-demographic features of households. Based on this multi-agent model, Aghababaei and Koliou [34] further added business agents and hospital agent, and similarly, they simulated the fired-hired process of employees in business agents and patient handling process using discrete event simulation in the hospital agent considering the disruption and recovery of infrastructure. Correspondingly, the job-hunting and injury treatment decision rule of household agents, as well as their interaction with other agents are designed to comprehensively estimate the number of affected employees and businesses. In addition, Hassan and Mahmoud [31] utilized an agent-based model to simulate the functional processes within hospitals and schools, and designed their decision-making heuristics to maximize functionality under various disruption conditions, such as using alternative staff, reducing patient treatment time, using hospital backup, and facilitating student admission/transfer.

### 3.3.2 Agent-based modeling for individual well-being impact

In terms of agent-based modeling for individuals' well-being impact of infrastructure disruptions, the simulation scale and emphasis are different for estimating objective and subjective well-being impact, though the people's adaptive or response behaviors are all incorporated similarly.

Compared to simulating the subjective well-being impact of disruptions, objective well-being impact focused on a relatively large scale, e.g., individual's changes of housing, food, working, and other daily life, without modeling the mechanism of individual's emotion and cognition. For example, Costa et al. [92] focused on the housing service of people under disruptive earthquake and designed the decision rules for household agents' temporary displacements and permanent relocations using flow chart and heuristic algorithm, respectively. At the same time, the decision algorithms of household agents take account of household socioeconomic demographics, social networks, and disaster preparedness. Crooks and Wise [93] built a spatially agent-based model to simulate people's survival considering the government's humanitarian assistance under disasters, and designed decision and interaction rules for two types of agents: Food distribution center agent and Individual agents. The individual agents' decision rule is driven by their survival needs, in the sense that individuals seek food in centers to increase their body energy. The travel behavior of individuals and their interaction collectively affect the performance of transportation. Agent-based modeling is widely used to incorporate individuals' travel activities to simulate road systems and several large simulation tools have been developed, such as MATSim, ALBATROSS, TRANSIMS[94]. Han et al. [95] applied MATSim to disaster scenarios and evaluated the impact of a disrupted road network due to storm surges on residents' travel activities, including working, shopping, schooling, leisure, and others.

As for subjective well-being impact estimation using agent-based modeling, in addition to individuals' adaptive behavior, their cognitions or emotions towards infrastructure disruptions are modeled by several methods, such as the empirical model, cognitive models, and dynamic modeling. Esmalian et al. [91] built a multi-agent model incorporating hazard agents, infrastructure agents, and household agents, to evaluate the impact of power outages on society's well-being, which is based on the hardship experience method. Households' tolerance level and hardship status are simulated by adopting empirical statistical models and setting decision processes. With a similar multi-agent framework, Yang et al. [96] further improved the decision rule of household agents to explore the negative well-being impact of disruptions (water, power, and transportation) and effective countermeasures. They proposed heuristic algorithms to estimate people's achievement of activities and intolerant states (societal impact) with limited resources (water and food) by minimizing suffering level, which is based on the deprivation cost method. To estimate the available resources, they designed the decision process of households conducting protective behaviors, such as going to stores and shelters for supplies. Silverman et al. [97] explored the population well-being impact of different healthcare interventions by building a three-level (individual, organization, and society) agent-based model. In the individual agent, a cognitive model (called PMFserv), including Motives, States, and Actions in appraisal loops, is developed to capture the mechanism of individuals' well-being impact. Individual agent's action decision is driven by satisfying their current state in a way that is consistent with their motives, and the current physiological, mental, and socioeconomic states of individuals contribute to the well-being impact. Valinejad et al. [98] developed a multi-agent-based stochastic dynamical model to estimate the mental and physical well-being impact of power outages and built an emotion (fear) dynamic model to explore mental well-being changes by considering the variation of risk perception, information-seeking behavior, flexibility, cooperation, and experience of individuals.

In general, agent-based modeling for societal impact estimation provides a powerful framework to incorporate the adaptive behaviors and interactions of multiple agents under infrastructure disruptions. This approach can improve the rationality of societal impact estimation, as well as facilitate exploring the mitigation effectiveness of different policies/measures. The key aspect of this approach is designing the decision rule for various agents, among which individual/household agents and institution agents are the most critical ones. Several methods have been integrated to support designing agent's decision rule, such as state chart, discrete event simulation, empirical model, and heuristic algorithm. It is worth noting that the simulation scale of an individual agent can be further narrowed deep into cognition level, in that sense, the cognitive model and emotion dynamic model can be utilized to capture the mechanism of human subjective well-being impact. Agent-based approaches have the advantages of developing a comprehensive agent-based model that incorporates multi scale and multi agent to simulate social systems. However, due to the flexibility and comprehension of the model, this type of method has the following shortcomings: 1) it is challenging to calibrate and validate the developed agent-based model of social systems, because the model usually includes many influencing coupled factors, which requires large or multi-source data to calibrate variables. The simulation results of social well-being impact are not easy to justify, and the measurement of the well-being of society is still an undergone question in social science. 2) the simulation results depend on the collective decisions and interactions of multiple agents, which are usually simplified and assumed using various methods without theoretical foundation, and a small inappropriate rule could induce different results. Addressing these challenges requires the cross-valuation of different data sources and integration with different study methods, for example, theoretical study, empirical study, and mathematical modeling. This future work will be further discussed in Section 4.

### 3.4 Big data-driven approaches

Big data-driven approaches explore and quantify the societal impact of infrastructure disruptions using the posts data from social media or mobility data from cell phones. The social consequences of disruptive events could be influenced by various coupled factors and challenging to capture, but with advancements and applications of contemporary information technologies and networked communication, human's actual activities and behaviors during disaster scenarios can be directly recorded, facilitating the analysis of societal impact patterns [99]. Specifically, this type of approach is specialized in understanding the reality of the dynamic human mobility across spatial-temporal scales and addressing the diverse needs of people under disruptive events. Integrated with the demographic characteristics of the affected population, social inequalities of the disaster impact can also be reflected. Furthermore, utilizing the large volumes of data generated from people, quantitative models can be established to sense the changes in human mobility (activities) and mental well-beings, and further help decision-makers to implement dynamic disaster risk reduction decision-making.

#### 3.4.1 Societal impact sensing by social media

Researchers have recognized the importance of social media within disaster management, and made efforts to acquire disaster situational awareness information by labeling disaster-related posts, geo-mapping the post, and conducting sentiment analysis. Distinguishing posts related to disasters from irrelevant posts is the first step for retrieving timely situational information from social media data. Disaster-related hashtags are commonly used to filter related posts, and various labeling taxonomies based on supervised learning are more informative, which can identify different types of damages (affected individuals, infrastructure, etc.). By aggregating disaster-related posts from the temporal dimension or mapping them from the spatial dimension, we can capture the patterns of individuals' posting activities and their correlations with disaster intensities or damages [100]. It is worth noting that spatial mapping requires location information, but only around 1%–4% of social media (e.g., Twitter) data posts are geo-tagged [101]. To mitigate this drawback, geoparsing (or geo-tagging) methods are developed to predict the locations of social media posts based on the content of the posts and the users' social network information [102]. Finally, sentiment analysis focuses on exploring people's sentiments, attitudes, emotions, and opinions about hazard events and facts, which are extracted from post contents by different supervised machine-learning approaches, including bag-of-words, part-of-speech tagging, n-grams, and keywords representing different sentiments [22]. Indeed, sentiment analysis can directly reflect residents' experiences and hardships in facing disruptions, and can therefore capture the nature and extent of societal impacts [103].

Using sentiment analysis of social media to investigate the individuals' mental well-being impact during disasters has been extensively studied over the past years. However, how to apply this technique to assess the disruptions of infrastructure and their impact on well-being (experienced hardship) has not yet been realized, which is also recognized as one of important future directions by Zhang et al. [22]. One category of sentiment analysis studies is labeling the posts with positive, neutral, or negative sentiments[104], while the other category refined the negative sentiment into fear, anger, and others [105]. These sentiments are usually extracted from posts in social media where the languages have been analyzed by machine learning (ML) or natural language processing (NLP) techniques[106]. For example, Li et al. [107] analyzed emotions and psychological states extracted from the datasets of Weibo users using linguistic inquiry, and word count (LIWC). Valinejad et al. [106] measured community well-being impact (social well-being and mental well-being) by the use of frequency of well-being-related words in tweets during a COVID-19 period using machine learning and text-mining tools (LIWC). These studies showed how the thoughtful application of simple NLP methods can provide insights into specific mental disorders and health under disasters. In recent years, several studies made attempts to apply sentiment analysis of social media to infrastructure disruption. For example, Roy et al. [108] presented a multilabel classification approach to identify the cooccurrence of multiple types of infrastructure disruptions considering the sentiment toward a disruption—whether a post is reporting an actual disruption (negative), or a disruption in general (neutral), or not affected by a disruption (positive). Zhang et al. [103] proposed a semi-automated social media analytics approach for Social Sensing of Disaster Impacts and Societal Considerations (SocialDISC), which enabled analysts to quickly capture emotional well-being impact (societal impact) associated with infrastructure disruptions from residents' reaction posts in social media. They focused only on the six basic emotions: anger, fear, surprise, sadness, joy, and disgust, and quantified the emotion score using the emotional lexicon collected and curated by the National Research Council of Canada [109].

In general, individuals' posts about disasters on social media could provide valuable and rich information about the descriptions of and people's reactions to disruption events, which could support a timely assessment of societal impacts, especially for the mental well-being impact by sentiment analysis. Compared with traditional questionnaire surveys (empirical approach), the social media approach could capture and analyze the time-sensitive societal impact information in a timely enough manner without conducting time-consuming and money-consuming social investigation. Especially, residents' memory may fade after the disruption passes, which limits the effectiveness of post-disaster survey, but social media could record the most real-time responses and reactions of people at the moment of disruptions. While the social media approach has lots of strengths, this approach still faces the following challenges: 1) whether social media users are a representative sample of the residents to reflect the whole well-being impact of the society is still not verified. As we can expect, the young are more active to post in social media than the old. 2) the disparity impacts among different social groups are difficult to investigate due to the data privacy issue. How to connect the socio-demographic information with the posting users is the key challenge. 3) identifying the location of the post is crucial in examining the spatial heterogeneity of the impact, while existing geo-parsing techniques have limitations in terms of the level of detail and level of accuracy for disaster situational information retrieval tasks.

### 3.4.2 Societal impact estimation by mobility data

Human mobility data usually record temporal and spatial information of human activities in a very detailed manner, allowing researchers to estimate people's daily movements and lifestyle patterns, especially their changes under disruptive events. Broadly speaking, human mobility data not only refer to the call detail records and global positioning system (GPS) data collected from smartphones, but also include other types of location-based data, such as the data from subway smart card, credit card transaction data, etc [99]. These datasets have been widely applied to solve urban challenges, such as population density estimation, dynamic traffic flow prediction, resource allocation, and modeling the spread of epidemics [110,111]. The applications of mobility data to the fields of urban resilience and disaster management are relatively limited, and they received more attentions in recent years. Several studies have used mobility data to analyze people's activity patterns before and after disasters [110,112] Actually, the mobility of a community is a complex but important variable for well-being [113], which could be holistically captured by the fluctuations of mobility data. For example, if households are economically impacted by disasters, or if they cannot access businesses (social institutions) due to road disruptions, or if institutions are closed due to damage, collective effects of these perturbations are reflected in changes in human activity patterns. Therefore, mobility data analysis could provide an integrative measure for examining the impacts of disruptive events.

Based on the mobility data, the societal impacts of the disaster are usually indicated by individuals' activity patterns and statistically calculated by the change percentage of individuals' POI visits or credit card transactions (CCT) to/in social institutions under disruptive events. For example, Podesta et al. [111] used the digital trace data related to unique visits to POIs in Houston during 2017 Hurricane Harvey, to quantify the community impact, which is measured by the percentage drop of POI visits (compared to its corresponding baseline over past three weeks). The POIs are divided into four groups according to their functions supporting people's activities: POIs essential for: 1) emergency preparedness, 2) emergency response, 3) lifestyle and well-being, and 4) recovery activity. Focused on the same disaster, Hong et al. [114] utilized large-scale smartphone geolocation data to quantify the community impact by the change percentage of people's mobility activity before and after the disaster as well. In addition to comparing to the baseline visits, Yabe et al. [115] quantified the business impact (e.g., grocery stores, hospitals, hotels, restaurant, supermarket etc.) by the difference between the observed daily visits under disaster and predicted daily visits under counterfactual situations (what if the disaster did not occur?), which are predicted by Bayesian structural time series model. Furthermore, CCTs data could be used to quantify the societal impact of disruptive events. For instance, Yuan et al. [116] quantified the community impacts by the maximum drop of CCT fluctuations of each sector (e.g., grocery store, drugstore, healthcare, etc.) in 2017 Hurricane Harvey, and they examined spatial patterns of disaster impacts by Moran I and gaussian regression analysis. Similarly, Dong et al. [117] measured the impact of a series of social protests on consumer actions (the number of customers) and personal consumption (the median spending) based on the ten million CCT data.

Using mobility data analysis, substantial existing studies have found that the societal impact of collective disruptions is not consistent across different spatial regions and different socioeconomic groups. These unequal impacts across spatial regions are relatively easier to identify using spatial statistics methods (e.g., Moran I), because mobility data contain sufficient location information. Due to the anonymity of the mobility data, it is difficult to connect the socioeconomic characteristics of individuals with their digital trace, consequently, leading to challenges in analyzing the disparate impact across various social groups. To solve this problem, Hong et al. [114] assigned each ping location from an individual device to the corresponding neighborhood grid cell based on its location, and each grid contains socio-demographic characteristics. As such, the activity pattern among groups with different socioeconomic status can be separately analyzed. They categorized these grids into 4 neighborhood groups based on disaster response and recovery patterns by an agglomerative clustering algorithm, and found clear socioeconomic and racial disparities in resilience capacity and evacuation patterns. This method aggregates people's movement into one grid, and it cannot capture their detailed activity patterns, such as visiting stores, healthcare, shelters, etc. To overcome this weakness, using location-based data, Esmalian et al. [118] built a population-facility network structure and dynamic clustering techniques to uncover the disparate access to grocery stores for socially vulnerable populations. They highlighted that disaster disproportionately exacerbated access disruptions to stores for socially vulnerable groups in the context of Hurricane Harvey. Overall, most existing studies focused on understanding and examining the unequal impact of disruptive events using mobility data, while few analytical methods and tools are available to guide mitigation measures in achieving equality and resilience goals. Fan et al. [119] attempted to calibrate models using 30 million anonymized smartphone-location data to optimize the distribution of facilities (stores), which is driven by minimizing the total travel distances of the residential populations to facilities and maximizing the equality of access to facilities.

In general, mobility data like POI visits could provide a holistic view of people's daily activity impact due to disruptive events as it captures population impacts, social institution interruptions, and infrastructure disruptions together. Compared with social media data, mobility data contain sufficient location and movement information of people but lack information related to individuals' opinions, perceptions, sentiments related to disasters. Thus, Mobility data analysis is suitable to measure the performance of social institutions and the objective well-being impact of individuals from the perspective of mobility activity, and social media data analysis advanced in capturing individuals' emotional well-being impact. In addition, mobility data analyses have the following shortcomings: 1) the location-intelligence data may not be representative of an affected population; in detail, mobile phone and credit card usage are lower in certain populations such as children, the elderly, the poor, and women [99]. 2) the baseline of mobility pattern is usually assigned by pre-disaster conditions of activities (normalcy), and external factors not related to the disaster impact would influence baseline mobility patterns, such as major community events or celebrations; resulting in increasing the bias of impact estimation. 3) Large-scale disasters may interrupt the power supply or destroy mobile towers, resulting in a complete loss of functionality of mobile phone networks, which may also cause data bias. 4) Existing mobility data analyses focus on understanding the unequal impact of disasters by statistics and machine learning methods, however, very few mathematical models are developed to conduct scenario analysis: predict the societal impact or mitigate the unequal impact.

# 4. Discussion

Section 3 reviews different approaches on modeling societal impact of infrastructure disruption. This section first compares different approaches with several criteria, and then summarizes the research challenge and future directions.

### 4.1 Comparisons of approaches

There exist several comparison criteria in the literature to review different modeling approaches. For example, Ouyang [2] compared the interdependent modeling approaches by the quantity and accessibility of input data, types of interdependencies, computation complexity, maturity, and resilience. Given our focus is on modeling the societal impact of infrastructure disruption, this paper includes the following three criteria to compare and discuss different approaches: 1) Quantity and accessibility of input data; 2) Types of societal impact; 3) Application scope of approaches.

**(1) Quantity and accessibility of input data**

The quantity of input data for different approaches is categorized into three levels: small, medium, and large amount of required input data. Also, considering the difficulty of data acquisition, this paper ranks the accessibility of input data by three levels: easy, medium, and difficult access of required input data. Based on these criteria for the input data, this paper compared those approaches introduced in Section 3, as shown in **Table 2**. Overall, the barriers to conducting big data-driven approaches are the most challenging, as the quantity of required data is large but difficult to access. In particular, majority of mobility data and social media data are recorded by apps in smartphones, which involve millions of users' opinions or location data across time and space. Due to privacy and confidentiality issues, these users' data are usually not allowed to be shared with the public, and only through research collaboration or high data collection costs. Also, scholars need to de-identify the mobility data to conduct such studies. In contrast, empirical approaches, which are usually involved with social surveys to collect relevant data, have the smallest barrier to conduct compared with others. In addition, although the quantity of input data in extended physical infrastructure modeling approaches is at a medium level, its accessibility is very difficult. To extend infrastructure modeling to capture its affected population (societal impact), the spatial distributions of infrastructure disruption usually need to be modeled using detailed information about components or characteristics of infrastructure, which are difficult to obtain due to privacy and national security issues. Finally, Agent-based approaches focus on modeling individuals' behaviors and interactions, which usually required large volume and multiple types of data to calibrate the parameters and validate the model.

**(2) Types of societal impact**

As introduced in Section 2.2, the societal impact of infrastructure disruption can be categorized into three types: social institution impact, objective well-being impact and subjective well-being impact. According to the classification of societal impact, this paper compared the applicability of four approaches, as shown in **Table 2**. Agent-based approaches can capture all three types of societal impact due to the flexibility of this approach to model interdependencies of infrastructure systems and social systems[2,120]. It is worth noting that when narrowing down the modeling scale, agent-based modeling could simulate the cognition process of an individual to further estimate the subjective well-being impact under disruptions. Big data-driven approaches mainly capture individuals' well-being impact because the data source is directly from human activities. Specifically, social media-driven and mobility data-driven analysis are used for sensing subjective and objective well-being impacts, respectively. Additionally, the extended physical infrastructure approaches focused on modeling the functionality of social institutions and affected individuals under disruptions. It is capable of evaluating the social institution impact and individual objective well-being impact, but it cannot capture people's subjective opinions about disruptions. This is a contrary situation for empirical approaches because they focusing on collecting people's opinions or feelings through questionnaires and are mainly used for capturing individuals' subjective well-being impact.

**(3) Application scope of approaches**

By reviewing substantial literature, this paper summarized the application scope of the four approaches into three groups: understanding the impact, scenario analysis, and social sensing. Empirically-based approaches and big data-driven approaches are mainly applied to understand the societal impact caused by disruptions. In detail, they focus on proposing quantitative instruments to indicate societal impact, and then using collected data to capture the societal impact pattern, such as identifying the main influencing factors, finding out the influencing pathway, and examining inequality of impact across spatial regions and social groups. Also, these two approaches are rarely used for scenario analysis independently, and in most situations, they are combined with other models to estimate/predict the societal impact of disruptive events. Extended infrastructure modeling and Agent-based modeling are popular in conducting scenario analysis and focused on establishing the relationship between disruptions and social systems (institutions and individuals). These two approaches are capable of estimating societal impact according to the intensity of hazards or extent of infrastructure disruption, and examining the effectiveness of different countermeasures. However, they have difficulties in capturing the inequity and disparity of the impact, which require fine-grained modeling and incorporating the heterogeneity of individuals, and agent-based models have potential to overcome this challenge. Finally, benefit from collecting real-time data, big data-driven approaches can be utilized to sense the societal impact information in a short-time manner after disasters.

**Table 2** Approach comparison from three criteria

| | Quantity of input data | Accessibility of input data | Societal impact types | Application scope |
|---|---|---|---|---|
| **1. Extended infrastructure modeling** | Medium | Difficult | Social institution impact;<br>Objective well-being impact | Scenario analysis |
| **2. Empirical approach** | Medium | Easy | Subjective well-being impact | Understanding the impact |
| **3. Agent-based approach** | Large | Medium | Social institution impact<br>Well-being impact | Scenario analysis |
| **4. Big data-driven approach** | Large | Difficult | Well-being impact | Understanding the impact；<br>Social sensing |

## 4.2 Challenges and future directions

Based on the review and approach comparison in Section 4.1, this subsection analyses the challenges and future directions of research in modeling societal impacts in disasters, as follows:

### (1) The measurement of societal impact

Scholars from different backgrounds proposed various instruments to indicate the societal impact of infrastructure disruptions, like the functionality reduction of social institutions, objective well-being impact (e.g., the number of individuals without food, water, housing, healthcare), subjective well-being impact (e.g., hardship experience, suffering level, negative emotion). There is a dearth of quantitative methods for quantifying the social costs of infrastructure disruptions and integrating them into infrastructure resilience assessments. In particular, in economic analyses of infrastructure resilience investments, the limited consideration and quantification of societal impacts would lead to underestimating the benefits of resilience investments and infeasibility of resilience investments. Future studies should aim for specifying empirical and quantitative methods for societal impacts/costs of various infrastructure services to complement the existing subjective measures.

Also, there is no general theoretical framework to support the societal impact measurement due to the multi-facet of society and the different purposes of studies, especially for the individual's well-impact measurement. So far, the capabilities approach may be a widely accepted theory to support the measurement of societal impact. Based on this theory, the hardship experience or suffering level of individuals is developed to better understand the negative impact, and this is regarded as the future direction of this field. However, the relevant existing studies mainly concentrated on the empirical study of the U.S., and future work could be extended to conduct more empirical case studies in other countries to further improve the theoretical and practical foundations of societal impact measurement. In addition, it is necessary to explore the methodology of modeling the mechanism of individuals' negative emotion, suffering level, or well-being impact in future work, e.g., individuals' cognition modeling using agent-based modeling[121], dynamics model, machine learning method, deep learning method. In fact, the ultimate goal of the future work is constructing a universally recognized theoretical framework and computational instruments for the societal impact of disruptions.

**(2) Model integration and co-simulation**

Different approaches have their own limitations and strengthens, and utilize only one approach usually cannot achieve accurate and entire assessment of the societal impact, especially for the application of scenario analysis. It is necessary to take full advantage of different approaches and integrate them to improve the estimation of societal impact. Empirical approaches advance in identifying the influencing factors and pathways of societal impact, which could facilitate building the relationship between disruptions and social systems. Extended infrastructure modeling is more suitable to derive the spatial distribution of service disruption from the perspective of individuals' or institutions' impact, thus, it can provide more accurate service disruptions for societal impact estimation. Agent-based approaches are very flexible and capable of simulating the decision-making processes of multiple agents (individuals and institutions) by mathematical equations or rules. Thus, the Agent-based model is suitable to be a unified framework that integrates all approaches; at the same time, agent-based simulation requires a large quantity of data to calibrate some parameters, which can be supplemented by other approaches. For example, big data-driven approaches collected plenty of real-time data, which can capture the real opinion about disruptions from social media, and the collective daily activity impact of individuals from smartphones. According to the purpose of the research, different approaches are encouraged to be combined to improve the accuracy of the societal impact estimation, e.g., the capabilities approach and big data-driven approaches [113].

**(3) Cross-validation of models**

It is crucial to validate the results of approaches before putting them into practical applications, while both the measurement and modeling of societal impacts are all involved with uncertainties, which increase the difficulties in approach validation. Also, attributing to the background or data accessibility restriction of scholars, existing studies usually validate the result by other literature indirectly or by one data source. However, considering the large uncertainties of societal impact estimation, it is necessary to conduct a cross-validation process with multiple datasets to enhance the developed model. Future work can be extended to combine empirical data from social surveys with the mobility data from smartphones under the same case study to validate the measurement of societal impact. At the same time, these data can be applied to cross-validate the results derived from extended infrastructure modeling and agent-based modeling.

**(4) Decision tools for mitigating inequity and societal impact**

Substantial existing studies revealed the inequity in societal impact of infrastructure disruptions, and highlighted that socio-economic vulnerable groups suffered disproportionally under the same extent of disruptions. Yet there are few analytic approaches to optimize decision-makers' measures to reduce or mitigate the inequity of the impact. In fact, the extended infrastructure approaches and agent-based approaches are suitable for conducting scenario analysis, which mainly includes examining the effectiveness of various countermeasures on mitigating societal impact. However, the countermeasures of inequity are rarely explored because the socio-economic statuses of individuals are difficult to be incorporated into modeling, and their influences on individual's autonomous decision-making increases the modeling difficulties. It is recommended to apply agent-based modeling frameworks to overcome these barriers by population synthetics, HUA, decision rule designing, and coupled with multiple data to validate.

## 5. Concluding remarks

Infrastructure disruption due to disasters could cause tremendous socio-economic impacts, in the past decades, substantial emphases were put on modeling interdependent infrastructure systems to better protect them and improve their resilience. As the role of infrastructure system in societal functioning has become increasingly critical, in recent years, scholars have gradually shifted their focus to study on understanding and modeling societal impacts of disruptions, and substantial progress has been made. To better comprehend the progress in current literature, this paper summarized the definitions and types of societal impacts of infrastructure disruptions, and reviewed quantitative studies about the measurement of societal impact, as well as their modeling approaches in the literature. The societal impact modeling approaches are grouped into four types: extended physical infrastructure modeling approaches, empirical approaches, agent-based approaches, and big data-driven approaches. For each type of the approach, this paper organizes relevant literature in terms of certain principles, such as the modeling idea, advantages, disadvantages, and the application scope.

In Section 4, different approaches are systematically compared and discussed according to three criteria: the quantity and accessibility of input data, types of societal impact, and application scopes. These comparisons facilitate scholars in understanding the characteristics, pros, and cons of each approach and then selecting appropriate approaches for their research. Building upon these, Section 4.2 outlines the remaining challenges and future directions in societal impact estimation, including the measurement of societal impact, model integration, cross-validation, and decision-making support tools. By improving the understanding of societal impact quantification progress in the existing literature, this review could provide an introduction to new scholars interested in this field and facilitate the development of these modeling approaches in disaster risk reduction, and further promote the resilient cities and society.

## Acknowledgements

This work is supported by National Natural Science Foundation of China [Grant Number: 72304039], and Beijing Normal University Start-up Projects of Scientific Research [Grand Number: 310432108].

## Declaration of Interests

The authors declare that they have no known competing financial interests or personal relationships that could have appeared to influence the work reported in this paper.